\documentclass[%
reprint,
 amsmath,amssymb,
 aps,prd
]{revtex4-1}

\usepackage{graphicx}
\usepackage{dcolumn}
\usepackage{bm}
\usepackage{epstopdf}

\newcounter{ichi}
\newcounter{ni}
\makeatletter
\@addtoreset{footnote}{page}
\makeatother
 
\begin{document}


\title{New Class of High-Energy Transients from Crashes of Supernova Ejecta with Massive Circumstellar Material Shells}

\author{Kohta Murase$^{1,2}$, Todd A. Thompson$^{1,3}$, Brian C. Lacki$^{1,3}$, and John F. Beacom$^{1,2,3}$}
\affiliation{
$^{1}$CCAPP, OSU, 191 W. Woodruff Ave., Columbus, Ohio 43210, USA\\
$^{2}$Department of Physics, OSU, 191 W. Woodruff Ave., Columbus, Ohio 43210, USA\\
$^{3}$Department of Astronomy, OSU, 191 W. Woodruff Ave., Columbus, Ohio 43210, USA
}

\date{July 12, 2011}
 
\begin{abstract}
A new class of core-collapse supernovae (SNe) has been discovered in recent years by optical/infrared surveys; these SNe suggest the presence of one or more extremely dense ($\sim {10}^{5-11}~{\rm cm}^{-3}$) shells of circumstellar material (CSM) on ${10}^{2-4}$~AU scales. We consider the collisions of the SN ejecta with these massive CSM shells as potential cosmic-ray (CR) accelerators. 
If $\sim 10$\% of the SN energy goes into CRs, multi-TeV neutrinos and/or GeV-TeV gamma rays 
almost simultaneous with the optical/infrared light curves are detectable for SNe at $\lesssim 20-30$~Mpc. 
A new type of coordinated multi-messenger search for such transients of duration $\sim 1-10$~months is required; these may give important clues to the physical origin of such SNe and to CR acceleration mechanisms.
\end{abstract}



\maketitle


\section{\label{sec:level1}Introduction}
The much-anticipated era of multi-messenger astronomy is coming. GeV-TeV gamma rays, which are powerful tracers of cosmic rays (CRs), are detected by \textit{Fermi} and ground-based Cherenkov detectors. Neutrinos, which can uniquely identify the production of hadronic CRs and probe dense sources, are detectable by the nearly-completed IceCube and the planned KM3Net~\cite{Ahr+04}. 
They are useful for studying sources especially when photons cannot escape directly.  

Violent explosions of massive stars, such as supernovae (SNe) and gamma-ray bursts (GRBs) may be prodigious neutrino/gamma-ray sources~\cite{HH02}. High-energy observations should reveal their physical origin, nonthermal processes, and extreme environments. 
TeV-PeV neutrino detections from extragalactic sources may be possible for GRBs~\cite{WB97,Mur08} and certain other kinds of SNe, e.g., those with hidden relativistic jets~\cite{RMW04}, relativistic pulsar winds~\cite{GS87}, or semi-relativistic external shocks~\cite{Wan+07}. But are SNe with more ordinary explosions also detectable?

In recent years, blind surveys for optical transients have discovered ultra-bright SNe such as SN 2008am~\cite{Cha+11}, 2008iy~\cite{Mil+10}, 2006gy~\cite{Ofe+07}, 2005ap~\cite{Qui+07} and 2003ma~\cite{Res+09}, which are much more luminous than ordinary SNe. Their physical origin is not settled, but a plausible interpretation is strong shock dissipation by collision with a massive ($M_{\rm sh}\sim 1-30 M_{\odot}$) circumstellar material (CSM) shell at $R \sim {10}^{2-3}$~AU~\cite{SM07,WBH07}, which is more or less analogous to the mechanism of type IIn SNe, though some may be pair-instability SNe~\cite{Gal+09}. 
The local rate of these ultra-bright SNe may be $\sim {10}^{3-4}$ times larger than the local apparent rate of classical long GRBs~\cite{Mil+09}. But SNe with CSM do not have to be optically ultra-bright, and the existence of massive CSM shells may be even more common. The existence of dense CSM shells is indicated from other SNe, e.g., SN 2006jc, 2005ip and PTF 09UJ~\cite{Imm+08,Smi+09}. 
Also, massive CSM eruptions were observed in luminous blue variables such as $\eta$ Carinae~\cite{Smi+03}, and are supported by the discovery of a self-obscured SN~\cite{Koz+10}.  

These observations motivate us to investigate the system of SN ejecta crashing into massive CSM shells at various radii. We consider the possibility of CR acceleration at strong shocks formed by the crashes, and suggest a new class of high-energy transients that are neither bursts like GRBs nor persistent sources like SN remnants. 
Given their rate and timescales, new types of searches at intermediate time scales, with coordinated observations of these extragalactic SNe at $\lesssim 100$~Mpc, are required. 
We adopt the conventional notation $Q_x = Q/{10}^x$. 

\section{The Scenario}
As the SN ejecta crashes into an external medium with density $n_{\rm ext}$, a pair of shocks (forward and reverse) are typically formed. As deceleration starts, significant fractions of the ejecta kinetic energy ($\mathcal{E}_{\rm ej} \sim {10}^{51}$~erg) are converted to the internal and kinetic energy of the shocked shells. The forward shock (FS) velocity in the deceleration phase is estimated to be $V_f \simeq 4000~{\rm km~s^{-1}}~\mathcal{E}_{\rm ej,51}^{1/2} n_{\rm ext,11}^{-1/2} R_{15.5}^{-3/2}$ in the non-radiative limit, and the reverse shock (RS) velocity can be estimated from $\rho_{\rm ej} V_r^2 \approx \rho_{\rm ext} V_f^2$, where $\rho$ is the mass density and $V_s$ is the shock velocity (where the index ``s'' indicates a forward or a reverse shock).  In the case of ordinary SN shocks sweeping up the interstellar medium, deceleration becomes significant around the Sedov radius, $R_d \approx R_{\rm Sed} = {(\frac{3 M_{\rm ej}}{4 \pi \rho_{\rm ext}})}^{1/3} \simeq 4.4 \times {10}^5~{\rm AU}~{(M_{\rm ej}/M_{\odot})}^{1/3} n_{\rm ext}^{-1/3}$. 
If massive ($M_{\rm sh} \gtrsim M_{\rm ej}$) CSM shells are located at $R_{\rm sh} \sim {10}^{2-4}$~AU, with CSM shell densities of $n_{\rm sh} \sim {10}^{5-11}~{\rm cm}^{-3}$, significant deceleration and shock dissipation are instead expected at $R_d \sim R_{\rm sh}$, which is thought to be the case for the bright SNe we consider here~\cite{Mil+10,SM07,WBH07}. Then, the shock velocities are evaluated via the Rankine-Hugoniot relations, and the FS will show a decelerating behavior self-similarly. Such violent eruption of massive CSM shells may be challenging for theories of the progenitor star, but one of the plausible mechanisms is the pulsational pair instability, which might occur at the deaths of main-sequence stars in the mass range $\sim 95-130~M_{\odot}$~\cite{WBH07}. 
Ejection of smaller mass would be more common, and for smaller CSM mass, the SN shock dynamics is not as affected and the dissipated energy would be scaled down as $\propto M_{\rm sh}/(M_{\rm ej}+M_{\rm sh})$. 

In many astrophysical shocks, sizable fractions of the kinetic energy seem to go into magnetic and CR energies. SN remnants are believed to be CR accelerators, and gamma rays from Galactic remnants are indeed detected by \textit{Fermi} and air-Cherenkov detectors (e.g.,~\cite{HESS09}). 
It is expected that plasma and/or MHD instabilities can amplify not only the downstream field but also the upstream field up to $\varepsilon_B \equiv 2 U_B/\rho V_s^2 \sim {10}^{-3}-{10}^{-2}$ (e.g.,~\cite{Bel78}), where $U_B$ is the magnetic energy density. If Galactic CRs come from SN remnants, CRs should carry $\epsilon_{\rm cr} \sim 0.1$ of the explosion energy, which seems consistent with observations, $2 U_{\rm cr}/\rho V_s^2 \sim 0.1-1$~\cite{HESS09,Hel+09}. 
In addition, CR acceleration in more dense environments ($n \sim {10}^{8-9}~{\rm cm}^{-3}$) is suggested by detections of gamma rays from V407 Cyg~\cite{Abd+10} and possibly $\eta$ Carinae~\cite{Wal+10}, which may imply that $\sim 1-10$\% of the kinetic energy still goes to CRs.   
 
Analogously, we may expect such magnetic field amplification and particle acceleration for the SN-CSM system. Unless the Thomson depth is $\tau_T \gg 1-10$, one may expect collisionless shocks, as in SN shock breakout~\cite{WL01}. 
The ion temperature behind the shock would become 
\begin{equation}
T_i^s \sim 9.8 \times {10}^{4}~{\rm eV}~V_{s,4}^2, 
\end{equation}
where the ion-ion collision frequency is $\nu_{ii} \sim 4 \times {10}^{-4}~{\rm s}^{-1}~n_{10} T_{i,5}^{-3/2}$. Electrons would be heated by ions while cooled by the Compton process and bremsstrahlung emission.  When only the Coulomb collision is considered, the equilibrium electron temperature is estimated to be 
\begin{equation}
T_e \sim {10}^{4}~{\rm eV}~T_{i,5}^{2/5} T_{\gamma,0.5}^{-8/5} n_{10}^{2/5}
\end{equation}
over the length of $\sim \nu_{ie}^{-1} V_s$, where $\nu_{ii} \gtrsim \nu_{ie} \sim 3 \times {10}^{-4}~{\rm s}^{-1}~n_{10} T_{e,4}^{-3/2}$. Here, the photon temperature is $T_{\gamma}^s = {(U_{\gamma}^s/a)}^{1/4} \sim 4.2~{\rm eV}~n_{10}^{1/4} V_{s,4}^{1/2}$ in the radiation-dominated case. On the other hand, one may expect that electromagnetic instabilities, e.g., $\nu_{\rm EM} \sim 4 \times {10}^{4}~{\rm s}^{-1}~n_{10}^{1/2} V_{s,4}$, are faster than all relevant collision frequencies~\cite{WL01}, such as $\nu_{ii}$ and $\nu_{ie}$, so collisionless shocks would be formed. 

However, the shocks can be radiation-mediated when $R_{\rm sh}$ is sufficiently small ($n$ is sufficiently large) and $\tau_T \gg c/V_s$~\cite{Wea76}. For example, in the SN 2006gy-like case, we have 
\begin{equation}
\tau_T^{\rm sh} = n_{\rm sh} \sigma_T {\Delta R}_{\rm sh} \simeq 210~n_{\rm sh,11} {\Delta R}_{\rm sh,15.5}, 
\end{equation}
where the radiative deceleration length for the FS, $L_{\rm dec}^f \sim \frac{c}{n_{\rm sh} \sigma_T V_f} \simeq 1.4 \times {10}^{15}~{\rm cm}~n_{\rm sh, 11}^{-1} V_{f,3.5}^{-1}$, is smaller than the width, ${\Delta R}_{\rm sh}$. In this situation, the flow is decelerated by radiation in the upstream region rather than collisions or plasma instabilities, and CRs would not carry a significant energy fraction, though the details are uncertain. Hence, we here consider CRs only when the shocks are not radiation-mediated.  
 
The CSM parameters ($n_{\rm sh}$, $R_{\rm sh}$ and ${\Delta R}_{\rm sh}$) are likely to have a variety of values, including cases with $\tau_T \lesssim 1-10$, so that one may expect the situation where CRs are accelerated at shocks formed by the collision between the SN ejecta and the CSM shell. For demonstrative purposes, we suppose CSM shells of $M_{\rm sh} \sim 1-30~M_{\odot}$ at $R_{\rm sh} \sim {10}^{15-17}$~cm, which are motivated by the SN-CSM collision model for explaining ultra-bright SNe~\cite{SM07,WBH07}, and use two representative cases: 
Model A ($n_{\rm sh}={10}^{11}~{\rm cm}^{-3}$, $R_{\rm sh}={\Delta R}_{\rm sh}={10}^{15.5}$~cm) and Model B ($n_{\rm sh}={10}^{7.5}~{\rm cm}^{-3}$, $R_{\rm sh}={\Delta R}_{\rm sh}={10}^{16.5}$~cm).
As for the SN ejecta, for simplicity, we assume fast moving, uniform ejecta with $\mathcal{E}_{\rm ej}={10}^{51}$~erg and $V_{\rm ej} \sim {10}^{4}~{\rm km}~{\rm s}^{-1}$ (where one may expect that the ejecta is several solar masses, which is lower than the CSM mass~\cite{WBH07}). 
The corresponding shock velocities (in the deceleration phase) are $V_f \simeq {10}^{3.5}~{\rm km}~{\rm s}^{-1}$ and $V_r \simeq {10}^{4}~{\rm km}~{\rm s}^{-1}$ in Model A, and $V_f \simeq {10}^{3.7}~{\rm km}~{\rm s}^{-1}$ and $V_r \simeq {10}^{3.9}~{\rm km}~{\rm s}^{-1}$ in Model B. These FS velocities are also consistent with observed ones (e.g.,~\cite{Ofe+07}).  
Model A is close to the case considered for explaining SN 2006gy~\cite{SM07,WBH07}, whose radiation energy and peak luminosity was $\mathcal{E}_{\rm ph} \sim {10}^{51}$~erg and $L_{\rm ph} \sim {10}^{44}~{\rm erg}~{\rm s}^{-1}$ respectively. In our setup, we may assume shock acceleration of CRs, except for the FS, since $L_{\rm dec}^f < {\Delta R}_{\rm sh}$ there. 
In Model B, a larger collision radius is assumed, which may be more similar to the case of dimmer but longer-lasting ultra-bright SNe such as SN 2008iy~\cite{Mil+10}, which had $\mathcal{E}_{\rm ph} \sim {10}^{50}$~erg and $L_{\rm ph} \sim {10}^{42.5}~{\rm erg}~{\rm s}^{-1}$, and we may consider CR acceleration at both the shocks. 
Note that, although we consider the uniform CSM shell with $\Delta R_{\rm sh} = R_{\rm sh}$, our results are not much changed if the CSM shell (with $\Delta R_{\rm sh} = R_{\rm sh}$) has a wind-like density distribution.

\section{Production of Neutrinos and Gamma Rays}
\emph{CR acceleration.---} When collisionless shocks are indeed formed, they would accelerate charged particles to high energies via the Fermi acceleration mechanism, producing a power-law distribution, $dN_p/dE_p \propto E_p^{-q}$ where $q \sim 2$~\cite{BE87}. The acceleration time scale is written as $t_{\rm acc}^s = \eta \frac{E_p}{e B c}$, where $\eta \sim \frac{20}{3} \frac{c^2}{V_s^2}$ in the Bohm limit. 
We assume that the magnetic field grows to a fraction of the equipartition value, $B_f \simeq 26~{\rm G}~\varepsilon_{Bf,-2.5}^{1/2} n_{\rm sh,11}^{1/2} V_{f,3.5}$ ($B_r \simeq 24~{\rm G}~\varepsilon_{Br,-2.5}^{1/2} {(M_{\rm ej}/M_{\odot})}^{1/2} R_{15.5}^{-3/2} V_{r,4}$), from analogy with SN remnants. Then we have 
\begin{eqnarray}
t_{\rm acc}^f &\sim& 2.6 \times {10}^{4}~{\rm s}~E_{p,100~\rm TeV}~\varepsilon_{Bf,-2.5}^{-1/2} n_{\rm sh,11}^{-1/2} V_{f,3.5}^{-3} \\
t_{\rm acc}^r &\sim& 2.7 \times {10}^{3}~{\rm s}~E_{p,100~\rm TeV} \varepsilon_{Br,-2.5}^{-1/2} {(M_{\rm ej}/M_{\odot})}^{-1/2} R_{15.5}^{3/2} V_{r,4}^{-3} \nonumber, 
\end{eqnarray}
which are larger than collision time scales at the thermal energy. 

The maximum CR energy, $E_{p,s}^{\rm max}$, is determined by comparison between the acceleration time and the cooling and dynamical time scales as well as the confinement condition. High-energy CRs lose their energies via adiabatic losses, and $pp$ and $p \gamma$ inelastic scatterings.  
The CSM density is so large that the $pp$ reaction is efficient. When the $pp$ cooling time, $t_{pp} = \frac{1}{n \kappa_{pp} \sigma_{pp} c}$, limits $E_{p,s}^{\rm max}$, 
we have 
\begin{eqnarray}
E_{p,f}^{\rm max} &\simeq& 81~{\rm TeV}~\varepsilon_{Bf,-2.5}^{1/2}  n_{\rm sh,11}^{-1/2} V_{f,3.5}^3 \\ 
E_{p, r}^{\rm max} &\simeq& 8.6 \times {10}^3~{\rm TeV}~\varepsilon_{Br,-2.5}^{1/2} {(M_{\odot}/M_{\rm ej})}^{1/2} R_{15.5}^{3/2} V_{r,4}^3 \nonumber.
\end{eqnarray} 
Here $\sigma_{pp} \simeq {10}^{-25.5}~{\rm cm}^2$ is the $pp$ cross section and $\kappa_{pp} \simeq 0.5$ is the $pp$ inelasticity. 
At large $R$, the adiabatic cooling time $t_{\rm ad}$, comparable to the dynamical time $t_s \approx R/V_{s}$, may become more important. When they limit $E_{p,s}^{\rm max}$, one obtains \begin{eqnarray}
E_{p,f}^{\rm max} &\simeq& 3.9 \times {10}^4~{\rm TeV}~\varepsilon_{Bf,-2.5}^{1/2} n_{\rm sh,11}^{1/2} V_{f,3.5}^3 t_{f,7} \\
E_{p,r}^{\rm max} &\simeq& 1.2 \times {10}^{5}~{\rm TeV}~\varepsilon_{Br,-2.5}^{1/2}  {(M_{\rm ej}/M_{\odot})}^{1/2} R_{15.5}^{-3/2} V_{r,4}^3 t_{r, 6.5} \nonumber.
\end{eqnarray}

At sufficiently high energies, $p \gamma$ processes occur, as often expected in extragalactic sources such as GRBs~\cite{WB97,Mur08,RMW04}. Both the photomeson cooling time ($t_{p \gamma}$) and Bethe-Heitler cooling time ($t_{\rm BH}$) depend on the radiation field. 
The emission from SNe is the reprocessed emission of ultraviolet, x-ray, and gamma-ray photons. In type IIn SNe, the reprocessing is expected to occur in the dense CSM. Calculating the spectrum requires detailed numerical modeling including the radiative transfer~\cite{SM07,Mor+09}, and we do not treat the details of its production here. Just for simplicity, we adopt a black-body spectrum that is valid in the thermal equilibrium limit (though the thermalization would be incomplete especially in Model B). In the case of SN 2006gy, the CSM shell is opaque even for Compton scattering, where stronger thermalization is expected and radiation should diffuse out of the shell~\cite{SM07},  and the interior temperature is set by assuming that radiation with $\mathcal{E}_{\rm ph}$ fills a sphere with radius $R_{\rm sh}+{\Delta R}_{\rm sh}$. In Model B, where $\tau_{T} \sim 1 < c/V_s$, radiation can escape the shell in the shock crossing time. We instead estimate the radiation energy density to be $\approx \frac{L_{\rm ph}}{4 \pi {(R_{\rm sh}+{\Delta R}_{\rm sh})}^2 c}$, where $L_{\rm ph} \approx \mathcal{E}_{\rm ph} V_s/R_{\rm sh}$.
The typical energy of protons interacting with $T_{\gamma}$ photons is  $E_p^{p \gamma} \simeq 1.6 \times {10}^5~{\rm TeV}~T_{\gamma,0}^{-1}$ and $E_p^{\rm BH} \simeq 4.8 \times {10}^3~{\rm TeV}~T_{\gamma,0}^{-1}$, respectively~\cite{GS87}.  

We numerically evaluated $t_{pp}$, $t_{p \gamma}$ and $t_{\rm BH}$, and determined $E_{p,s}^{\rm max}$ by comparing $t_{\rm acc}$ to cooling time scales and $t_s$~\cite{Mur08}, where we assumed $\mathcal{E}_{\rm ph}={10}^{51}$~erg in Model A and $\mathcal{E}_{\rm ph}={10}^{50}$~erg in Model B. 
We obtained $E_{p,r}^{\rm max} \simeq 3.2 \times {10}^{3}$~TeV in Model A, and $E_{p,f}^{\rm max} \simeq 5.0 \times {10}^{3}$~TeV and $E_{p,r}^{\rm max} \simeq 2.0 \times {10}^{4}$~TeV in Model B. 
We found that both $t_{p \gamma}$ and $t_{\rm BH}$ are not very relevant in our cases. 

Accelerated CRs are mostly confined and produce mesons via inelastic $pp$ scattering, which leads to the production of neutrinos and gamma rays. At $R_{\rm sh}$, the efficiencies of the $pp$ reaction during $t_{s}$ are estimated to be~\cite{Mur08,RMW04}
\begin{eqnarray}
f_{pp}^{\rm sh} &\approx& t_{f}/t_{pp}^{\rm sh} \simeq470 ~{R}_{\rm sh, 15.5} n_{\rm sh,11} V_{f,3.5}^{-1} \\
f_{pp}^{\rm ej} &\approx& t_{r}/t_{pp}^{\rm ej}  \simeq 13~R_{\rm sh, 15.5}^{-2}  (M_{\rm ej}/M_{\odot}) V_{r,4}^{-1} \nonumber.
\end{eqnarray}
Hence, the $pp$ reaction should be efficient for typical parameters, $R_{\rm sh} \lesssim {10}^{16.5}$~cm and $M_{\rm sh} \sim 1-30~M_{\odot}$, which are consistent with the parameters suggested for explaining ultra-bright SNe in the SN-CSM collision model~\cite{Mil+10,SM07,WBH07}. (In ordinary SN remnants, $f_{pp} \sim {10}^{-5} {(M_{\rm ej}/M_{\odot})}^{1/3} n_{\rm ext}^{2/3} V_{f,3.5}^{-1}$ at the Sedov time.) 
Although CRs may make further pions through diffusion even after the shock crossing, CRs would also have adiabatic cooling due to an expansion with $\sim M_{\rm ej} V_{\rm ej}/(M_{\rm ej}+M_{\rm sh})$ (which is the fluid velocity of the merged ejecta). Our estimates on the $pp$ efficiencies are relatively conservative and adequate. 

\emph{Neutrinos.---} To calculate neutrino and gamma-ray spectra from the $pp$ reaction, we performed numerical calculations~\cite{KAB06}, where the CR spectrum was assumed to be $d N_p/d E_p \propto E_p^{-2} {\rm e}^{-(E_p/E_{p,s}^{\rm max})}$, normalized by the total CR energy, $\mathcal{E}_{\rm cr} \equiv \epsilon_{\rm cr} \mathcal{E}_{\rm ej}$.   
The effects of radiative and/or hadronic cooling of mesons on spectra, which are relevant in GRBs~\cite{Mur08,RMW04} and magnetars~\cite{GS87}, can be neglected. 

\begin{figure}
\includegraphics[width=0.99\linewidth]{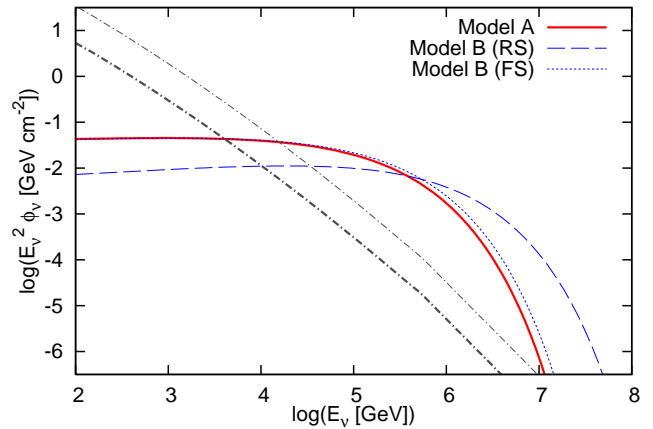}
\caption{
\small{Energy fluences of muon neutrinos from a SN crashing into dense CSM, where $\varepsilon_B={10}^{-2.5}$, $\epsilon_{\rm cr}=0.1$ and $d=10$~Mpc are assumed. 
Thick and thin curves represent Model A and Model B, respectively (see text). 
The dotted-dashed curves show the zenith-angle-averaged ANB within a circle of radius $1^\circ$~\cite{Abb+10}; we use $\Delta t={10}^7$~s for Model A (thick line) and $\Delta t={10}^{7.8}$~s for Model B (thin line).}}
\end{figure}
After flavor mixing, $\sim 1/6$ of the pion energy is carried by each neutrino flavor, so the muon neutrino fluence is 
\begin{equation}
E_{\nu}^2 \phi_{\nu} \sim 6 \times  {10}^{-2}~{\rm GeV}~{\rm cm}^{-2}~{\rm min}[1,f_{pp}] \epsilon_{\rm cr,-1} \mathcal{E}_{\rm ej,51} d_1^{-2},
\end{equation}
which agrees with our numerical results shown in Fig.~1. Here $d=10~{\rm Mpc}~d_1$ is the distance.
The signals compete with the atmospheric neutrino background (ANB), so one has to know both the source direction and timing for detections. The SN direction can be determined well by optical/infrared observations. Importantly, the high-energy transients from the new SN class discussed here have \textit{much longer} duration (months-years) than GRBs (seconds-minutes)~\cite{WB97,Mur08} and SNe with relativistic components (minutes-days)~\cite{RMW04,GS87}, and thus form a new type of neutrino sources. 
In the shock model for ultra-bright SNe such as SN 2006gy~\cite{SM07}, thermal photons leave the source when the photon diffusion time is comparable to the shell expansion time, where
\begin{equation}
t_{\gamma-D} \approx \frac{{({\Delta R}_{\rm sh})}^2}{2 c} n_{\rm sh} \sigma_T \sim {10}^{7}~{\rm s}~n_{\rm sh,11}^{-1} V_{f,3.5}^{-2}
\end{equation}
(which is consistent with the observation, $\mathcal{E}_{\rm ph} \sim {10}^{51}$~erg and $L_{\rm ph} \sim {10}^{44}~{\rm erg} {\rm s}^{-1}$). 
For the neutrino search by IceCube-like detectors, we have to set a time window $\Delta t$, which is relevant to estimate the ANB. In Model A, it would be appropriate to use $\Delta t={10}^7$~s since the duration of the SN thermal emission is $t_{\gamma-D} \sim t_f \sim {10}^7$~s, where the muon yield from SN-CSM neutrinos for IceCube is $N_{\mu,>4~{\rm TeV}} \sim 2$. In an optically thin case like Model B, the SN emission time is order of 
\begin{equation}
t_{s} \simeq 5.0 \times {10}^{7}~{\rm s}~R_{\rm sh,16.5} V_{s,3.8}^{-1}
\end{equation}
(which is consistent with $\mathcal{E}_{\rm ph} \sim {10}^{50}$~erg and $L_{\rm ph} \sim {10}^{42.5}~{\rm erg} {\rm s}^{-1}$), and we obtain $N_{\mu, >20~{\rm TeV}} \sim 1$ for the FS ($N_{\mu, >50~{\rm TeV}} \sim 0.2$ for the RS) for this time window. For up-going neutrino sources, attenuation in Earth should be considered, but will be modest at a wide range of zenith angles for the most important energies~\cite{DA81}.  

The rate of SNe with dense and massive CSM is uncertain, but a few~\% of all SNe may be such systems~\cite{Mil+09,Koz+10,Smi+09}, so that their rate within 20~Mpc is order of $\sim 0.1~{\rm yr}^{-1}$. 
Note that the cumulative background muon neutrino flux, $E_{\nu}^2 \Phi_{\nu} \sim 2.7 \times 10^{-9}~{\rm{GeV~cm^{-2}~s^{-1}~sr^{-1}}}$, though comparable to that from GRBs~\cite{WB97,Mur08}, is less than the ANB up to $E_\nu \sim 300$~TeV, so that we focus on detections of individual nearby explosions. 

\begin{figure}
\includegraphics[width=0.99\linewidth]{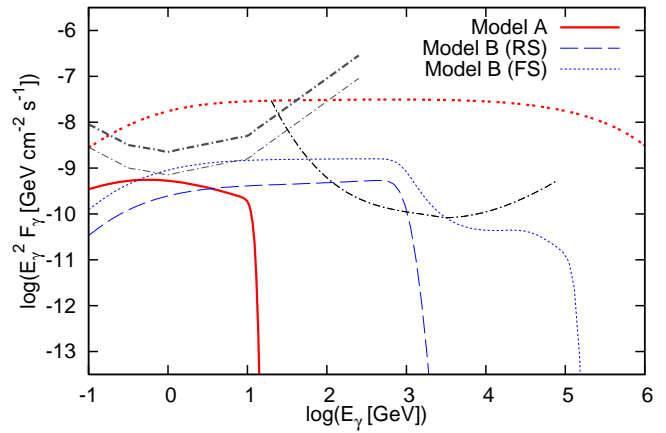}
\caption{
\small{
Energy fluxes of pionic gamma rays, corresponding to Fig.~1. 
Gamma-ray attenuation inside and outside the source is included; the double-dotted curve shows the intrinsic spectrum without attenuation in Model A. 
Left dot-dashed curves show \textit{Fermi}/LAT differential sensitivities at $t={10}^{6.5}$~s ($\sim t_{r}$ in Model A) and $t={10}^{7.5}$~s ($\sim t_{r}$ in Model B). The 100~hr differential sensitivity of CTA is also overlaid (right).} 
 }
\end{figure}

\emph{Gamma rays.---} Neutral pions lead to gamma rays that are interesting targets for \textit{Fermi} and future Cherenkov telescopes such as Cherenkov Telescope Array (CTA), where one has to consider their interactions inside and outside the source. At relevant energies, there will be attenuation on matter (Bethe-Heitler pair-creation) and radiation ($\gamma \gamma$ pair-creation) in the source. They are also attenuated by the extragalactic background light (EBL).

In Fig.~2, the attenuated pionic gamma-ray fluxes are shown, taking into account gamma-ray attenuation numerically. For demonstrative purposes, the non-attenuated flux in Model A is also shown. Here, for simplicity, we employ $\sim \exp(-\tau_{\gamma \gamma}-\tau_{\rm BH})$ for the screen region and $\sim 1/(1+\tau_{\gamma \gamma}+\tau_{\rm BH})$ for the emission region. 
For example, the Bethe-Heitler and $\gamma \gamma$ pair-creation depths in the CSM shell are estimated to be $\tau_{\rm BH}^{\rm sh} \simeq 3.2~n_{\rm sh,11} {\Delta R}_{\rm sh,15.5}$ at $\sim$~GeV and $\tau_{\gamma \gamma}^{\rm sh} \simeq 3000~T_{\gamma,0}^{3} {\Delta R}_{\rm sh,15.5}$ at $\sim 260~{\rm GeV}~T_{\gamma,0}^{-1}$, respectively. (Note that the photomeson and photodisintegration processes can also happen at sufficiently high energies. When $\tau_T$ is sufficiently small, low-frequency synchrotron far-infrared emission may also increase the attenuation far above a TeV). Outside the source, the EBL attenuation is significant only at $\gtrsim 100$~TeV for $d \sim 10$~Mpc.

In Model A, the Bethe-Heitler and $\gamma \gamma$ attenuation would make it difficult to detect $\sim$~GeV and $\sim$~TeV gamma rays, respectively, although the attenuated flux just represents the relatively conservative case (see below). Also, though we show the differential sensitivities of \textit{Fermi} and CTA, the integrated sensitivities over several energy bins are much better, which would help detection of the signal. In Model B, $\tau_{\rm BH}^{\rm sh} \ll 1$ and the $\gamma \gamma$ attenuation is negligible at $\lesssim$~TeV, so that gamma rays seem detectable by \textit{Fermi} for $d \lesssim 20-30$~Mpc, which motivates searches for $\sim 0.1-1$~yr transients via multi-year \textit{Fermi} observations. With coordinated follow-up searches, $\sim 0.1-1$~TeV gamma rays may also be detected by Cherenkov telescopes such as CTA. 
The High Altitude Water Cherenkov Experiment (HAWC), with a larger field of view and lower sensitivity, may also be helpful for nearby SNe.  

The neutrino signature is quite direct and more important as a smoking gun of the CR acceleration, while the gamma-ray signature would be more complicated.  
While we are here mainly concerned with pionic gamma-ray emission that is the more direct hadronic signal, gamma rays are also produced by electrons, which come from muon-decay, primary acceleration, and pair-creation processes. When they lose energy via inverse-Compton (and synchrotron) processes, electromagnetic cascades can be induced, so that gamma-ray signals may be enhanced. 
Let us estimate the cascade effects on the gamma-ray emission briefly. As noted above, gamma rays of $\sim 260~{\rm GeV}~T_{\gamma,0}^{-1}$ lead to generating energetic pairs. The synchrotron and inverse-Compton cooling times are $t_{e-{\rm syn}} \simeq 7.7 \times {10}^{3}~{\rm s}~\gamma_{e,3}^{-1} B_1^{-2}$ and $t_{e-{\rm IC}} \simeq 2.2 \times {10}^{2}~{\rm s}~\gamma_{e,3}^{-1} T_{\gamma,0}^{-4}$, respectively (while the bremsstrahlung cooling time is $t_{e-{\rm brem}} \simeq 1.0 \times {10}^4~{\rm s}~n_{\rm sh,11}^{-1} {(\ln \gamma_{e,3})}^{-1}$). 
If interactions with matter could be neglected, the situation would be much simpler. When we consider only a one-zone black-body radiation field (though it may not be a good approximation), unless the synchrotron cooling becomes more important, we obtain an inverse-Compton cascade, which leads to a broken power-law energy spectrum with a high-energy cutoff $E_{\gamma}^{\rm cut}$ (e.g., \cite{CA97}). 

In Model A, because of the large $\gamma \gamma$ pair-creation depth, a flat energy spectrum is expected from $\sim 0.1~{\rm GeV}~(E_{\gamma}^{\rm cut}/10~{\rm GeV})^2 T_{\gamma,0}$ to $E_{\gamma}^{\rm cut} \sim 10$~GeV (and the photon index becomes $\sim 1.5$ at lower energies). The cascaded flux is $\sim 2 \times {10}^{-8}~{\rm GeV}~{\rm cm}^{-2}~{\rm s}^{-1}$ (which is comparable to the intrinsic flux without attenuation), enhancing the $\sim 1-10$~GeV flux only by a modest factor; this estimate was also checked via the numerical cascade calculation~\cite{MB10}. 
However, interactions with matter cannot be neglected, so that even $\sim$~GeV gamma rays are attenuated mainly because of the Bethe-Heitler attenuation in the CSM shell and the eventual gamma-ray emission is expected at even lower energies, $\sim 10~{\rm keV}-100~{\rm MeV}$, below the range of our interest (and note that the radiation temperature around the photosphere would be lower). Then, further downgrades are unavoidable, since the Compton opacity becomes larger than unity for photons, and electrons mainly lose their energies via the Coulomb interaction ($t_{e-{\rm Coul}} \simeq 1.2 \times {10}^{3}~{\rm s}~\gamma_{e,2} n_{e,11}^{-1}$), and the bremsstrahlung process can overtake the inverse-Compton process. Then, the GeV-TeV emission would be significantly suppressed as expected above, and low-energy emission is largely thermalized. 
However, the CSM might be clumpy as suggested in SN 2005ip~\cite{Smi+09}, where the Bethe-Heitler process might not be so significant depending on the unknown volume filling factor. If that is true, \textit{Fermi} may find the signal if $\sim 3$\% of gamma rays (including both the attenuated/cascaded components) can escape from the source, where the flat energy spectrum in the $\sim 0.1-10$~GeV range is expected because of the inverse-Compton cascade. But, since strong thermalization is required in the SN 2006gy-like case, this is recognized as optimistic for detectability. 

On the other hand, in Model B, the $\gamma \gamma$ pair creation is relevant only at $\gtrsim$~TeV and the Bethe-Heitler process is negligible, so that the attenuated flux shown in Fig.~2 (which is essentially the intrinsic flux without attenuation) gives the dominant contribution for the detectability of \textit{Fermi} and CTA. Here, the inverse-Compton cascade is suppressed by the synchrotron cooling ($t_{e-{\rm syn}} < t_{e-{\rm IC}}$) but the secondary synchrotron flux in the $\sim 10~{\rm keV}-100~{\rm MeV}$ range (see also the next section) would be enhanced by a factor. 
Note that the $\gamma \gamma$ pair-creation opacity by synchrotron x rays coming from the generated secondary pairs is small enough. Let us suppose all of the pair luminosity goes into x rays of $E_{X} \sim$~keV, though it is not true. The maximum number of x ray photons is $\approx \frac{L_e}{4 \pi {(R_{\rm sh}+{\Delta R}_{\rm sh})}^{2} c E_X}$ and a keV photon travels $\sim 3.8 \times {10}^{15}~{\rm cm}~n_{\rm H,7.5}^{-1}$ in a homogeneous medium, so that the $\gamma \gamma$ pair-creation opacity is $\sim 6.2 \times {10}^{-5}~L_{e,41.5} R_{\rm 16.5}^{-2} E_{X,\rm keV}^{-1} n_{\rm H,7.5}^{-1}$. 

Generally speaking, calculating the detailed gamma-ray spectrum is not easy without knowing the multi-zone spectral energy distribution of target photons, which depends on details of the model via the frequency dependence of the opacity and the density gradient in the CSM~\cite{Mor+09}. The radiation deviated from a black-body spectrum, including x-ray emission from thermal bremsstrahlung expected in the downstream region, may affect results. Though we defer such studies, we have discussed both the relatively conservative (with attenuation) and optimistic (with cascade) cases, which are enough for our purpose of demonstrating detection potential of the gamma-ray signal. 
Also, one should keep in mind that the cascade outside the source is also initiated, but the resulting gamma-ray emission cannot be detected unless the intergalactic magnetic field is weak enough~\cite{MB10}.

\section{Implications and discussions}
We have shown that SNe crashing into dense CSM are interesting targets for current and near-future high-energy detectors. Importantly, new types of coordinated multi-messenger searches are required to detect such $0.1-1$~yr transients.

CR acceleration in dense surroundings is also motivated by recent gamma-ray observations~\cite{Abd+10,Wal+10}. Both detections and non-detections of SN-CSM emission will be useful,  since physical mechanisms in such extreme environments are uncertain, especially when $\tau_T \gtrsim 1$. In particular, neutrinos have the benefit of probing hadronic CR accelerators, and they can escape earlier than thermal photons, which may be delayed by diffusion.  
In addition, the detections of signals would support the SN-CSM scenario~\cite{SM07,WBH07} rather than the pair-instability scenario~\cite{Gal+09}, useful for revealing the origin of bright SNe. 
  
One may expect synchrotron emission in the infrared-to-gamma-ray bands, as electrons of energy $E_e$ emit photons with $\sim 44~{\rm keV}~E_{e,\rm TeV}^2 B$. But the emission should be reprocessed to energies $\lesssim 10$~keV and/or strongly thermalized when $\tau_T \gg 1$, so we basically expect thermal emission observed from such SNe. 
But the synchrotron emission can be seen in the sufficiently hard x-ray range, if the collision happens at $\tau_T < 1$, as in Model B. The unabsorbed energy flux is estimated to be $\sim 7 \times {10}^{-13}~{\rm erg}~{\rm cm}^{-2}~{\rm s}^{-1}~f_{\rm syn}  {\rm min}[1,f_{pp}] \epsilon_{\rm cr,-1} \mathcal{E}_{\rm ej,51} d_1^{-2} t_{s,7.8}^{-1}$, where $f_{\rm syn} \leq 1$ is the efficiency of synchrotron cooling. In Model B, the synchrotron cooling is more important than the synchrotron self-Compton and external Compton cooling at sufficiently high energies, so that pionic gamma rays are dominant in the GeV-TeV range while the synchrotron component, whose high-energy photon index is $(q+2)/2 \sim 2$, is relevant below $\sim 100$~MeV. (But other cooling processes such as the Coulomb interaction becomes more important especially at $E_e \lesssim 0.1-1$~GeV.) Hence, this signal will be an interesting target for the near-future x-ray monitor \textit{NuStar} (softer x rays can be masked by strong thermal bremsstrahlung emission that seems to be observed~\cite{Mil+10}).
The radio emission is suppressed by the Razin effect, free-free absorption, and synchrotron self-absorption when the collision radius is small enough, though it may be expected at very large radii. 

For more quantitative theoretical studies, hydrodynamical simulations with radiation transfer and CR back-reaction are desirable. But our results are enough for the purpose of this work, which, in part, is to motivate new searches starting now. 
The relevant quantities, $\mathcal{E}_{\rm cr}$, $E_{p,s}^{\rm max}$ and $q$, have uncertainties, but we could see a source up to $d \sim 30-60$~Mpc if $\epsilon_{\rm cr}$ is larger than 0.1. 
The spectral index also affects the results. Although we adopt $q=2$, steeper indices lead to lower muon yields from SN-CSM neutrinos. But harder indices may also be expected since the gas may be radiation-dominated and/or the shock may be CR-mediated~\cite{BE87}. 
Since $E_{p,s}^{\rm max}$ depends on shock velocities, the non-uniformity of the ejecta affects the RS velocity and ratio of the RS dissipation to the FS dissipation~\cite{Che82}, but its pre-collision velocity distribution is uncertain since the ejecta may sweep the CSM before the collision. Also, the shock evolution may be radiative rather than adiabatic~\cite{Mar+10}.

Plasma effects can modify the results, via e.g., wave damping by neutral particles or radiation.   
Especially, $E_{p,s}^{\rm max}$ may be limited by the size of the ionization region, since damping occurs in a time $\sim \frac{1}{n_{n} <\sigma_{i-n} v_{\rm th}>} \sim {10}^{0.5}~{\rm s}~n_{n,7.5}^{-1} T_{0}^{-0.4}$ in the neutral region~\cite{KC71}. 
Although the CSM gas would be initially neutral, ionization in the downstream and upstream region (not far from the shock) seems expected observationally~\cite{Smi+09} and theoretically~\cite{HS84} since the post-shock temperature is high. 

We have considered extragalactic SNe with dense and massive CSM shells. Such collisions may happen even for GRBs, where the jet breakout emission can be expected. The high-energy neutrino and gamma-ray emission is also possible, which would be more or less analogous to the (sub-)photospheric emission~\cite{Mur08,RMW04}.  
In the Galaxy, $\eta$ Carinae is a promising candidate that showed violent mass eruptions~\cite{Smi+03}. The radius of its massive nebula is larger than our typical values that are required for bright SNe. But the neutrino detection seems possible if the star explodes, since $N_{\mu, > \rm TeV} \sim 2 \times {10}^5 \mathcal{E}_{\rm cr,50} (M_{\rm sh}/10 M_{\odot}) R_{\rm sh,17}^{-2} V_{s,3.5}^{-1}$ (where $f_{pp}<1$).
For smaller CSM mass, though the radiation might push the CSM shell, SN dynamics are not largely affected as in ordinary SNe~\cite{BK00}, where $\mathcal{E}_{\rm cr}$, $E_{p,s}^{\rm max}$ and $f_{pp}^{\rm sh}$ are much smaller. 
Detections would be challenging, but it may be interesting for a Galactic event. 

After this work was submitted and put onto the arXiv (arXiv:1012.2834), we became aware of Ref.~\cite{KSW11}, which is closely related to ours, and which supports our claims that these unusual SNe are interesting and that their high-energy emission is an important probe.
We thank K. Ioka, C. Kochanek, C. Rott and R. Yamazaki for discussions. 
This work is supported by JSPS and CCAPP (KM), E. C. Howald Presidental Fellowship (BCL), Sloan Fellowship and NSF Grant AST-0908816 (TAT), and NSF CAREER Grant PHY-0547102 (JFB). 


\end{document}